\documentclass[12pt, a4paper]{article}
\usepackage{amsmath, amsfonts}
\usepackage{cite}
\usepackage{amssymb}
\usepackage[dvips]{graphicx}
\usepackage{epsfig}
\begin{document}


\newtheorem{definition}{Definition}
\newtheorem{theorem}{Theorem}
\newtheorem{lemma}{Lemma}


\title{Improved Exponential Time Lower Bound of Knapsack Problem under BT model}

\author{Xin Li, Tian Liu, Han Peng, Hongtao Sun\footnote{Corresponding author, sunht\_pku@tcs.pku.edu.cn}, Jiaqi Zhu\\ \it School of EECS, Peking University, P.R.China}

\maketitle
\begin{abstract}
M.Alekhnovich et al. recently have proposed a model of algorithms,
called BT model, which covers Greedy, Backtrack and Simple Dynamic
Programming methods and can be further divided into fixed, adaptive
and fully adaptive three kinds, and have proved exponential time
lower bounds of exact and approximation algorithms under adaptive BT
model for Knapsack problem which are $\Omega(2^{n/2}/\sqrt
n)=\Omega(2^{0.5n}/\sqrt n)$ and
$\Omega((1/\epsilon)^{1/3.17})\approx\Omega((1/\epsilon)^{0.315})$(for
approximation ratio $1-\epsilon$) respectively (M. Alekhovich, A.
Borodin, J. Buresh-Oppenheim, R. Impagliazzo, A. Magen, and T.
Pitassi, Toward a Model for Backtracking and Dynamic Programming,
\emph{Proceedings of Twentieth Annual IEEE Conference on
Computational Complexity}, pp308-322, 2005). In this note, we
slightly improved their lower bounds to
$\Omega(2^{(2-\epsilon)n/3}/\sqrt{n})\approx　\Omega(2^{0.66n}/\sqrt{n})$
and
$\Omega((1/\epsilon)^{1/2.38})\approx\Omega((1/\epsilon)^{0.420})$,
and proposed as an open question what is the best achievable lower
bounds for knapsack under adaptive BT models.
  \end{abstract}

\section{introduction}
\label{chap1}
Many combinatorial optimization problems are
NP-complete and probably have no polynomial time
algorithms\cite{Garey1979}. It is presumed that there are only
exponential time algorithms for these problems, however, this has
not been proved. It is very hard to prove exponential time lower
bound for these problems under universal model of algorithms, unless
prove $P\neq NP$. Nevertheless, under some restricted model of
algorithms, it is possible to prove exponential lower bounds for
NP-complete problems. Alekhnovich et al have proposed a restricted
model of algorithms, called BT model, which covered greedy
algorithm, backtrack and simple dynamic programming
\cite{Alekhnovich2005}, and have proved exponential lower bound for
Knapsack, Vertex Cover and SAT under this model. In this note, we
slightly improved their lower bounds of exact and approximation
algorithms for Knapsack problem under the adaptive BT model.

\section{A Glimpse on BT model}
\label{chap2}

In this section we quickly recall the relevant notions from BT
models \cite{Alekhnovich2005}.

\subsection{A description of BT model}

A combinatorial optimization problem $P$ is represented by a set of
data items $D$ and a set of choices $H$. Each data item represents a
partial structure of an instance. The set of choices contains all
the possible choices which can be applied to data items. For
example, in Knapsack problem, each item can be a data item and
"chosen to be in the knapsack" and "chosen not to be in the
knapsack" can be the set of choices. In the following, $O(S)$
denotes the set of all the orderings of all the elements in $S$.

A BT algorithm $A$ to a problem $P$ is comprised of an ordering
function $r_A^k$ and a choice function $c_A^k$:
$$r_A^k: D^k\times H^k\mapsto O(D)$$ is the ordering of unprocessed
data items made by algorithm after the first $k$ data items has been
processed; $$c_A^k: D^{k+1}\times H^k\mapsto O(H\cup\{\perp\})$$
represents the constraints made by algorithm when it makes choice
for the $k+1th$ item according to the processed $k$ data items and
the choices for them. The algorithm consider only the choices before
$\perp$. If $r_A^k$ is a constant function and does not depend on
$D^k$ or $H^k$, then it's called fixed. If $r_A^k$ depends on $D^k$
but not on $H^k$, then it's called adaptive. If $r_A^k$ depends on
both $D^k$ and $H^k$, then it's called fully adaptive. In this note,
only adaptive BT algorithm is considered.

\subsection{Lower bound strategies for BT model}

\begin{figure}[h]
\centering
\includegraphics[scale=0.6]{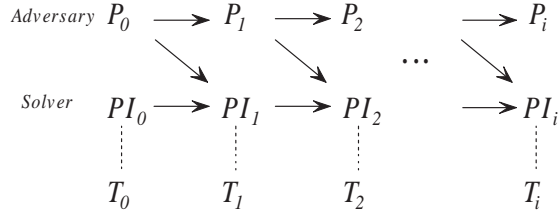}
\caption{Model of solver-adversary} \label{chap2:update:2_1}
\end{figure}

The lower bound strategies for BT model takes the form of a game
between the adversary and the solver(the BT
algorithm)\cite{Pudlak2000}. In the game, since at the beginning the
solver cannot see all the input data items, so the adversary always
tries to produce a difficult problem instance for the solver. The
game can be viewed as a series of rounds. The $i$th round is
composed of three parts: $P_{i}$, $PI_{i}$ and $T_{i}(i=0,1,2...)$.
$P_{i}$ is the finite set of data items owned by the adversary which
cannot be seen by the solver in the $i$th patter. $PI_{i}$ is the
set of data items representing a partial instance of the problem
which have been seen by the solver in the $i$th pattern. And $T_{i}$
is the set of partial solutions to $PI_{i}$. See figure
\ref{chap2:update:2_1}.

At the beginning of the game, the solver gives an ordering rule on
the input data items. And the adversary constructs rule of deleting
data items according to the solver's rule and gives $P_{0}$. In this
pattern, $PI_0$ is empty and $T_0$ is also empty. In the
$i$th$(i\geq1)$ pattern, solver picks a data item $a$ from $P_{i-1}$
and add it too $PI_{i-1}$, and gets $PI_{i}=PI_{i-1}\cup{a}$. Then
computes $T_i$ in which every solution is an extension of some
solution in $T_{i-1}$. Then adversary deletes $a$ from $P_{i-1}$ and
some other data items and gets $P_i$. This process continues until
$P_i$ is empty. In the last pattern, if $PI_i$ is not a valid
instance or $T_i$ contains the optimum solution of $PI_i$, then
solver wins, or else adversary wins.

If the set of all the solutions to $PI_i$ can be classified to some
equivalent classes and for any partial solution $PS$ to any
equivalent class, there exists an instance $A\subseteq PI_{i}\cup
P_{i}$ so that every optimum solution of $A$ contains $PS$, then PS
is called "indispensable". If there exists a pattern in which the
number of all the indispensable solutions is exponential with the
scale of the problem, then the exponential lower bound of the
problem can be achieved under the fixed or adaptive BT model.

\section{Improved lower bound of Knapsack}
\label{chap3}

In this section we give the improved exponential lower bounds of
exact algorithm and approximation algorithm on Knapsack under
adaptive BT model. This improvement is obtained by setting
parameters differently in original proofs in the work of
M.Alekhnovich et. al \cite{Alekhnovich2005}.

\noindent\textbf{Knapsack Problem}

\noindent Input: $n$ pairs of non-negative integers,
$(x_{1},p_{1}),...,(x_{n},p_{n})$ and a positive integer $N$.
$x_{i}$ represents the weight of the $i$th item and $p_{i}$
represents the value of the $i$th item. $N$ is the volume of the
knapsack.

\noindent Output: \begin{displaymath} \{i|max\{\sum_{i\in
S}p_{i}|S\subseteq {1,2,...,n},\sum_{i\in S}x_{i}\leq
N\}\}\end{displaymath}

\noindent\textbf{Simple Knapsack Problem}

\noindent Input: $n$ non-negative integers $\{x_{1},...,x_{n}\}$ and
a positive integer $N$. $x_{i}$ is the weight and value of the $i$th
item and $N$ is the volume of the knapsack.

\noindent Output: \begin{displaymath} \{i|max\{\sum_{i\in
S}x_{i}|S\subseteq [n],\sum_{i\in S}x_{i}\leq N\}\}\end{displaymath}

Simple Knapsack problem is also $NP$ complete\cite{Garey1979}. So it
is only needed to prove the exponential lower bound for simple
Knapsack. In the following, we denote a simple Knapsack problem with
$n$ items and knapsack volume of $N$ with $(n,N)$.

\subsection{Lower Bound of Exact Algorithm} \label{chap3:update}

M.Alekhnovich et. al proved the following theorem in
\cite{Alekhnovich2005}.

\noindent\textbf{Theorem 1.} For simple Knapsack problem $(n,N)$,
the time complexity of any adaptive BT algorithm is at least
$${{n/2}\choose {n/4}}=\Omega(2^{0.5n}/\sqrt n).$$

Next we prove the following theorem.

\noindent\textbf{Theorem 2.}For simple Knapsack problem $(n,N)$,
$\forall 0<\epsilon\leq1/2,\exists N_{0},s.t.$ when $N>N_{0}$, the
time complexity of any adaptive BT algorithm is at least
$${(2-\epsilon)n/3 \choose
(2-\epsilon)n/6}=\Omega(2^{(2-\epsilon)n/3}/\sqrt{n})\approx　\Omega(2^{0.66n}/\sqrt{n}).$$

\noindent\textbf{Proof:} Let $I=\{0,1,...,\frac {3N}{n}\}$, all the
weights of the items are from $I$.

This is an constructive proof containing three steps.

\noindent\textbf{Step 1.}

Solver chooses the first $(2-\epsilon)n/3$ items. After each,
adversary deletes some items from the remaining by the following two
rules. Let $S$ be the set of the items which have been seen by the
solver.

\noindent (1) If $S_{1},S_{2}\subseteq S$, then delete items with
weight of $|\sum S_{1}-\sum S_{2}|$;

\noindent (2) If $S_{1}\subseteq S$, then delete items with weight
of $|N-\sum S_{1}|$;

\noindent\textbf{Step 2.}

Let $P$ be the set of the first $(2-\epsilon)n/3$ items chosen by
solver. Now we prove: $\forall Q\subset P$ and
$|Q|=(2-\epsilon)n/6,\exists R\subset I-P,|R|=(1+\epsilon)n/3$ and
$\sum_{x\in Q}x+\sum_{x\in R}x=N$. See figure \ref{chap3:update3_3}:

\begin{figure}[h]
\centering
\includegraphics[scale=0.5]{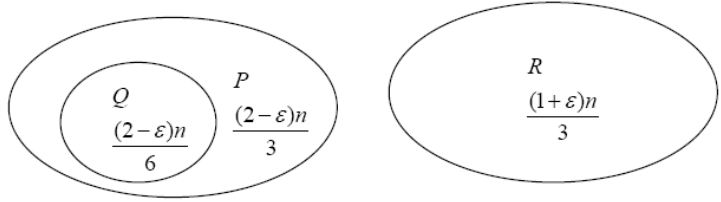}
\caption{Constructing an indispensable partial solution}
\label{chap3:update3_3}
\end{figure}

$\forall Q \subset P$, let
$U=3^{n},a=\frac{3}{(1+\epsilon)n}(N-\sum_{x\in Q}x),
J=\{i|i\in\mathbb{Z}\wedge i\in[a-U,a+U]\}$ (later we will prove $J
\subset I$). Adversary chooses $\frac{(1+\epsilon)n}{3}-2$ items in
$J$. After each, adversary will delete some items from the remaining
by the following two rules. Let $S$ be the set of items chosen by
solver and adversary.

\noindent (1) If $S_{1},S_{2}\subseteq S$, then delete items with
weight of $|\sum S_{1}-\sum S_{2}|$;

\noindent (2) If $S_{1}\subseteq S$, then delete items with weight
of $|N-\sum S_{1}|$;

Let the sum of the weighs of these $\frac{(1+\epsilon)n}{3}-2$ items
be $w$. Then the adversary can choose items with weight bigger or
smaller than $a$ so that $$|w-a(n/2-2)|\leq U.$$

\noindent\textbf{Step 3.}

Adversary choose a pair of items from the following $U+1$
pairs(later we will prove they are in $I$), such that the weight of
these two items and the weight of the previously chosen items sum to
$N$:
$$(\frac{v}{2}-i,\frac{v}{2}+i),i=1,2,...,U+1$$
in which $v=N-\sum Q-w$.

Since in the process of choosing items, the number of all the items
does not exceed $n$, let this number be $n_t$. Order these $n_t$
items by a fixed order, then every item corresponds to a single bit
of a $n_{t}$-bit 0-1-(-1) string. And every such string corresponds
to a deleted item in the previous process. In detail, let the sum of
the weights of items corresponding to 1 in the string be $s_1$, and
let the sum of the weights of items corresponding to -1 in the
string be $s_{-1}$. If $s_{1}-s_{-1}>0$, then delete item with
weight of $s_{1}-s_{-1}>0$; if $s_{1}=0$, then delete item with
weight of $N-s_{-1}>0$. So the number of the deleted items does not
exceed $U$. So in these $U+1$ pairs of items, there must be one pair
which is not deleted. Then we can choose this pair in the third
step.

Next we prove all the weights of the chosen items are in $I$.

Because $$a=\frac{3}{(1+\epsilon)n}(N-\sum Q)$$ and
$$0\leq \sum Q \leq
\frac{(2-\epsilon)n}{6}\cdot
\frac{3N}{n}=\frac{(2-\epsilon)3N}{6},$$ so
$$\frac{3\epsilon N}{2(1+\epsilon)n}\leq a \leq
\frac{3N}{(1+\epsilon)n},$$ clearly,
$$0\leq \frac{3\epsilon
N}{2(1+\epsilon)n}\leq a\leq
\frac{3N}{(1+\epsilon)n}\leq\frac{3N}{n}.$$

Since
$$|w-a(n/2-2)|\leq U,$$
$$a=\frac{3}{(1+\epsilon)n}(N-\sum
Q)$$ and
$$v=N-\sum Q-w,$$
so
$$|v-2a|\leq U,$$
that is
$$a-\frac{U}{2}\leq\frac{v}{2}\leq
a+\frac{U}{2}.$$ So in the third step, the possible lightest and
heaviest items are $a-\frac{U}{2}-U-1$ and $a+\frac{U}{2}+U+1$. So
in order that all the items chosen in the third step are from $I$,
it is only needed that

$$ \left\{
\begin{array}{rcl}
a-\frac{U}{2}-U-1 & \geq & 0 \\
a+\frac{U}{2}+U+1 & \leq & \frac{3N}{n} \\
\end{array}
\right.,$$ that is

$$ \left\{
\begin{array}{rcl}
N & \geq & \frac{3(1+\epsilon)n3^{n}+2(1+\epsilon)n}{3\epsilon} \\
N & \geq & \frac{3(1+\epsilon)n3^{n}+2(1+\epsilon)n}{6\epsilon} \\
\end{array}
\right.,$$ So it is only need that $N\geq
\lceil\frac{1+\epsilon}{\epsilon}\rceil n3^{n}$. In fact, $N\geq
\lceil\frac{1+\epsilon}{\epsilon}\rceil n3^{n}$ is also sufficient
to $J\subset I$.

Next we prove that solver must keep all $Q$ in $P$ as partial
solution in the computation tree, or else adversary can let the
solver cannot find the optimum solution.

We use the reduction to absurdity. Let $Q_{1},Q_{2}\subset P$ and
$|Q_{1}|=|Q_{2}|=(2-\epsilon)n/6$. $R_{1}$ and $R_{2}$ correspond to
$Q_{1}$ and $Q_{2}$ respectively. If solver does not keep $Q_1$,
then adversary deletes all the items except $R_{1}$. If solver keeps
$Q_2$ to get the optimum solution, then there are only three
cases(see figure \ref{chap3:update:3_4}):

\begin{figure}[h]
\centering
\includegraphics[scale=0.5]{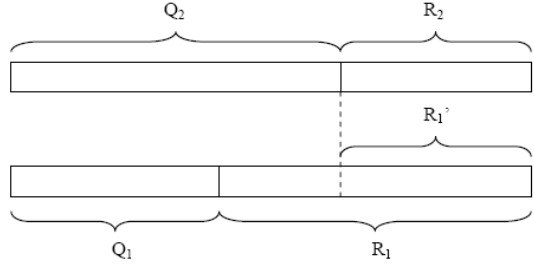}
\caption{Proof of indispensable of partial solution of Knapsack}
\label{chap3:update:3_4}
\end{figure}

\noindent 1) $\sum Q_{2}+\sum R_{1} <N$

The optimum solution $N$ cannot be achieved in this case.

\noindent 2) $\sum Q_{2}+\sum R_{1} =N$

So $\sum Q_{1}=\sum Q_{2}$. This is impossible. Since according to
the first rule in the first step, the last chosen item in the first
step which is contained in $Q_{1}$ or $Q_{2}$ but not both cannot be
chosen in the first step.

\noindent 3) $\sum Q_{2}+\sum R_{1}>N$, that is $\sum Q_{2}>\sum
Q_{1}$

Assume $R_{1}'$ is a subset of $R_{1}$ and it leads to the optimum
solution with $Q_{2}$. Then there are two cases:

\noindent 3.1) there is only one item in $R_{1}-R_{1}'$

According to the first rule in the first step, this item should be
deleted.

\noindent 3.2) there are two or more than two items in
$R_{1}-R_{1}'$

If there is only one item in $R_{1}-R_{1}'$ which is chosen in the
third step, then according to the first rule in the second step, it
should be deleted.

If there are two items which are chosen in the third step, then all
the items in $R_{1}'$ are chosen in the second step. Let the
choosing order of these items be $i_{1},...,i_{s}$, then according
to the second rule in the second step, the last item $i_{s}$ should
be deleted.

So every $Q$ in $P$ is indispensable. So the time complexity is at
least $${(2-\epsilon)n/3 \choose
(2-\epsilon)n/6}=\Omega(2^{(2-\epsilon)n/3}/\sqrt{n})\approx　\Omega(2^{0.66n}/\sqrt{n}).$$
\begin{flushright}
$\blacksquare$
\end{flushright}


\subsection{Lower bound of approximation algorithm}
\label{chap3:update}

M.Alekhnovich et. al proved the following theorem in
\cite{Alekhnovich2005}.

\noindent\textbf{Theorem 3.} For simple Knapsack problem, using
adaptive BT algorithm, to achieve $1-\epsilon$ approximation ratio,
the time complexity is at least $\Omega((1/\epsilon)^{1/3.17})$.

Next we prove the following theorem:

\noindent\textbf{Theorem 4.} For simple Knapsack problem, using
adaptive BT algorithm, to achieve $1-\epsilon$ approximation ratio,
the time complexity is at least $\Omega((1/\epsilon)^{1/2.38})$.

\noindent\textbf{Proof:}In theorem 2 we proved that for given
$n,N=\lceil\frac{1+\delta}{\delta}\rceil n3^{n},0<\delta<0.5$, the
time complexity of simple Knapsack problem is
$\Omega(2^{\frac{(2-\delta)}{3}n}/\sqrt{n})$. So the optimum
solution cannot be achieved by algorithm with complexity of
$\gamma=o(2^{\frac{(2-\delta)}{3}n}/\sqrt{n})$. Since the weight of
item is integer, so the upper bound of approximation ratio is
\[\frac{N-1}{N}\sim1-\frac{1}{\lceil\frac{1+\delta}{\delta}\rceil
n3^{n}}\sim1-\widetilde{O}({\gamma}^{-\frac{3}{2-\delta}\log_{2}3})\sim1-\epsilon,\]
so
\[\gamma=\Omega((1/\epsilon)^{\frac{2-\delta}{4.75}}).\]

For $n,N,0<\delta<0.5(N>\lceil\frac{1+\delta}{\delta}\rceil
n3^{n})$, let $N=\lceil\frac{1+\delta}{\delta}\rceil
n_{0}3^{n_{0}}$. So we can set the weights of $n-n_{0}$ items in the
$n$ items to $0$.
\begin{flushright}
$\blacksquare$
\end{flushright}

\section{Discussion}
\label{chap4}

In this paper, we slightly improved the exponential time lower
bounds of exact and approximation algorithms under adaptive BT model
for Knapsack problem which are
$\Omega(2^{(2-\epsilon)n/3}/\sqrt{n})\approx　\Omega(2^{0.66n}/\sqrt{n})$
and
$\Omega((1/\epsilon)^{1/2.38})\approx\Omega((1/\epsilon)^{0.420})$.

We do not know whether our lower bounds are optimal. An interesting
question is: what are the best achievable lower bounds for Knapsack
problem under adaptive BT model?

\end{document}